\documentstyle[12pt,epsfig,psfig]{article}

\newcommand{\be}{\begin{equation}}
\newcommand{\ee}{\end{equation}}

 1
\font\elevenrm=cmr10 scaled\magstep 1
 1

\def\reff{\hang\noindent}

\def\deg {^\circ}

\begin{document}
\vspace*{1.8cm}
  \centerline{\bf THE BEPPOSAX VIEW OF THE GALACTIC CENTER REGION}
\vspace{1cm}
  \centerline{SIDOLI LARA}
\vspace{1.4cm}
  \centerline{SSD-ESA/ESTEC}
  \centerline{\elevenrm Postbus 299, NL-2200 AG Noordwijk, The Netherlands
}
\vspace{3cm}
\begin{abstract}

The Galactic Center Region has been surveyed in 1997--1998 
with the  Narrow Field Instruments
on-board the BeppoSAX satellite (2--10~keV energy range).
The X--ray emission from Sgr~A*, the putative supermassive black hole
at the center of our Galaxy, has been measured and its spectrum studied,
together with the X--ray emission from several  
bright X--ray sources,  Low Mass X--ray Binaries (LMXBs)
containing neutron stars and black holes located
within the region ($|l|<2\deg)\times(|b|<1\deg$).
New point--like X--ray sources have also been discovered.
The results on the diffuse emission coming from the SgrA Complex are   
summarized here: it displays 
a  two--temperature  thermal spectrum (kT$_{\rm 1}\sim$ 0.6~keV  
and  kT$_{\rm 2}\sim$8~keV) and  an  energy--dependent morphology, with the 
hard emission (5--10 keV) elongated along the 
galactic plane and the soft one (2--5 keV) spatially correlated with 
the  radio halo of  SgrA~East.
The results of the study of the diffuse X--ray emission from other
three fields in the Galactic Center Region 
containing the giant molecular clouds Sgr~B2, Sgr~C and Sgr~D
are   reported for the first time.
The diffuse emission from the  direction of these 
three  molecular clouds
shows a double temperature nature as well,
with a prominent iron line at $\sim$6.7~keV, the intensity and equivalent width
of which depends on the galactic longitude. 
Moreover, the Sgr~B2 spectrum shows an intense
6.4~keV line of fluorescent origin. 
While the hard component of the Galactic Center diffuse emission displays a similar
temperature with respect to the Galactic Ridge, the low temperature plasma 
is significantly  softer. This may be due to a net contribution 
produced from the accretion of material onto old isolated neutron stars located 
inside the giant molecular clouds.

\end{abstract}
\vspace{2.0cm}

\section{ Introduction }

The Galactic Center (hereafter GC) Region is still one of the most 
enigmatic regions of   our Galaxy,
despite the deep investigations performed in the entire 
electromagnetic spectrum.
The extreme  interstellar absorption severely affects its
observability, especially in the 
optical/ultraviolet range. 
The interstellar medium is transparent  
for energies below 2$\mu$m and above $\sim$2 keV. 
Thus, the X--ray band is one of the privileged 
observing windows for the GC.
Since a large concentration of mass exists towards the GC 
direction, the crowding of sources, at all energies, 
is severe as well. 
For this reason,  
instruments with  a good spatial resolution are needed.

Several peculiar 
objects are present within a few parsec from the Center of our Galaxy,
including the SgrA Complex, a remarkable ensemble of structures
observed in the radio band. 
It is composed by at least four distinct objects (see Fig.~1): 
the SgrA East shell,  the SgrA East halo, 
the SgrA West HII region, with its mini--spiral shape, and the compact 
non--thermal radio source Sgr~A*.
The first two structures are   non--thermal radio sources.
Their real nature is not clear from the radio observations alone.  
They  might be supernova remnants
or  by--products of a past  outburst of activity from the GC.

\vspace{-0.7cm}
\begin{figure}[thb]
 \begin{center}
  \mbox{\epsfig{file=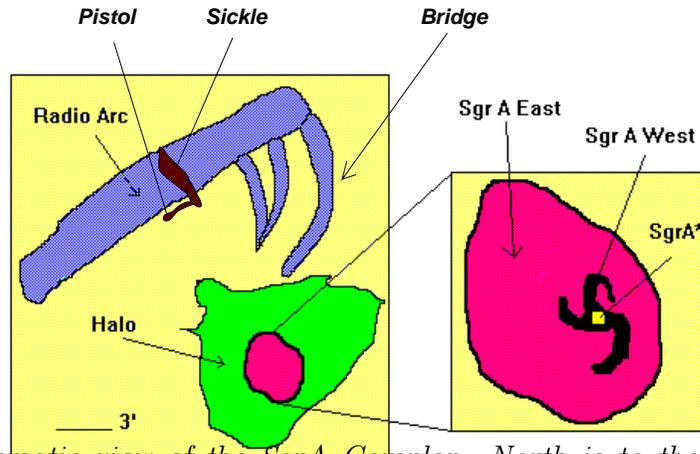,width=13cm}}
  \vspace{-5cm}
  \caption{\em {A schematic view of the SgrA Complex. North is to the top, 
East to the left. 1 arcmin corresponds to about 2.5 pc
at 8.5 kpc. All these structures resides within $\sim$20~pc from the GC, but their
relative positions are still poorly known. 
The triangular--shaped  halo is a distinct object
with respect to the SgrA~East shell (Pedlar et al. 1989), 
and it is probably located at the GC, or maybe in front of it.
Sgr~A* is embedded in the  ionized gases of SgrA~West, and the SgrA~East shell is located
behind it}}
\end{center}
 \end{figure}

Sgr~A*, located at the dynamical  center of our Galaxy,  is
thought to be a super--massive black hole of about 2.5$\times10^{6}$  
solar masses.
The presence of such a large dark mass  towards it is supported by
the measurements of  the mass distribution and 
dynamics of infrared stars   in the central parsec of the Galaxy (e.g., Ghez et al. 1998).
The overall spectrum of Sgr~A* is still poorly determined.
Up to now there are  no sicure counterparts of Sgr~A* 
at any other wavelengths, with the exception of 
the recent detection with $Chandra$ (Baganoff et al. 1999).

The Center of our Galaxy is rather silent at high energies, emitting
at a level which is several orders of magnitude below the Eddington luminosity
for a super--massive black hole of its mass.
In fact, the hard X--ray images obtained with  coded--mask detectors   
have shown that the
high energy activity previously detected with non--imaging instruments 
and   associated with the GC, 
is instead produced by two black hole candidates 
of stellar origin, 1E~1740.7--2942 and GRS~1758--258,  
unrelated to the GC.
Previous X--ray observations   led to the discovery that   
a high density of point--like sources 
exists towards the GC. 
These sources include several    transients, i.e. objects that are
bright in X--rays only sporadically.
It is  unclear whether this concentration follows 
the   stellar mass  density or if there is
evidence of  an additional   population of X--ray sources.
The bright sources (L$_{\rm X}> 10^{36}$ erg  s$^{-1}$) are    binaries
containing compact objects (neutron stars or black holes) 
accreting matter from stellar companions (probably of low mass).
Some of these sources are particularly interesting, belonging to the
class of ``micro--quasars", with  relativistic, 
double-sided    radio jets.

The GC region is also characterized by  strong diffuse
X--ray emission, first observed with the $Einstein$ Observatory (Watson et al. 1981).
The $Ginga$ satellite revealed the presence of a 6.7~keV line 
emission from the Galactic
Plane which was particularly intense towards the GC direction (Koyama et al. 1989).
The ASCA satellite confirmed the presence of this diffuse component, which  
 extends
symmetrically with 
respect to the GC. 
Its spectrum   is well described  by a thermal   hot plasma 
with a temperature of $\geq$7~keV, but there is also   evidence 
for a multi--temperature
plasma. In fact, several emission lines are present, with the K--lines from 
iron and sulfur (at $\sim$2.4 keV) particularly bright (Koyama et al. 1996).
The nature of the diffuse emission is still unknown, especially since   
its temperature is too high to allow the confinement of the emitting plasma
by the galactic gravity. 
Part of it (especially that at  the lower energy  
region of the spectrum) could be due to the   integrated 
thermal emission from supernova remnants. Other emitting processes
have been proposed in order to reject the thermal nature
of this component or to decrease its temperature: 
non--thermal emission from SNRs, inverse Compton 
scattering by relativistic
electrons,  emission lines of non--thermal origin
produced during capture of  electrons by accelerated ions, 
charge exchange by low energy 
heavy ions with neutral gas, non--thermal emission from the
interaction  of low energy cosmic ray electrons with the interstellar
medium (see, e.g., the recent results of Valinia et al. 2000). 

Another diffuse component was discovered with ASCA: a   6.4~keV 
iron line component  of fluorescent origin, the extent of which well
correlates    
with the distribution of the molecular clouds 
in the GC region (Koyama et al. 1996).
This emission is thought to be the result of irradiation of the 
dense molecular clouds by hard photons  coming from bright X--ray sources, 
located inside or outside the cloud.
However, this emission seems too intense  to be due to 
reprocessing of hard X--rays   from 
any known   source in the GC region. 
Thus,  a possible explanation is that the illuminating source 
could have been Sgr~A*, during a past phase of high-energy activity (Koyama et al. 1996; 
Churazov et al. 1996).
Thus, the study of the diffuse component can also cast light on the 
past high energy activity from the GC itself.

I report here on the results of the survey (1997-1998) of the GC
region performed with the Narrow Field Instruments (NFI) on-board 
the BeppoSAX satellite (Boella et al. 1997). Only   data obtained
 from the MECS instrument (2--10 keV energy range) are relevant here. 
In particular, the main objects of this study are the following: 

\begin{itemize}

\item
 the spectral properties
of the  point--like X--ray sources, low mass X-ray binaries (LMXBs)
 containing neutron stars and black holes, both displaying 
 persistent and transient emission; 
 
 \item 
 the X--ray counterpart of  Sgr~A*, its spectrum and its emission level;
 
 \item
 the spectrum and spatial distribution of the diffuse emission from the
 GC Region.
\end{itemize}

\section{Spectroscopy of the Point--like Sources}

\begin{figure}[!h]
 \begin{center}
  \mbox{\epsfig{file=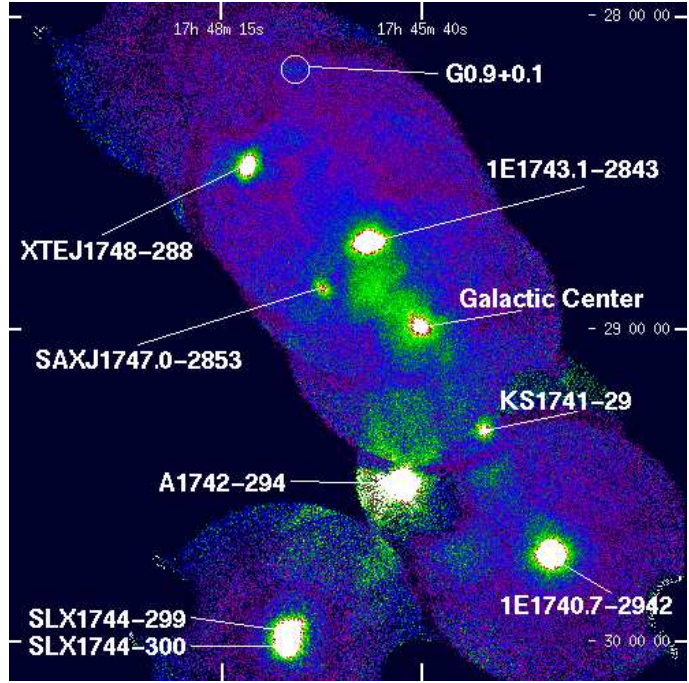,width=9cm}}
  \caption{\em {Mosaic of the MECS  pointings in the GC direction (2--10~keV). 
  Each circle represents the MECS FOV, with a 
  diameter of about 1 degree ($\sim$150 pc at the GC distance)}}
\label{fig:mosaic}
 \end{center}
\end{figure}

The region of the sky covered by the BeppoSAX NFI
observations is shown in Fig.~\ref{fig:mosaic}. 
A  pointing on the 
black hole candidate GRS~1758--258, $\sim4.5$ degrees away from
the GC, is not included in this mosaic 
(see Sidoli et al. 1999 and Sidoli 2000 for 
a detailed description of the analysis and results).
 
We discovered new X--ray sources: the  X--ray emission of plerionic 
origin coming from the radio core 
of the composite supernova remnant G0.9+01 
(Mereghetti et al. 1998; Sidoli et al. 2000) and weak X--ray emission from
a still unidentified object, discovered near the molecular cloud Sgr~D 
at the position  R.A.=$17^{{\rm h}}~48^{{\rm m}}~16^{{\rm s}}$,
Dec.=$-28\deg~08'~13''$ (J2000, $\sim$1$'$ uncertainty). Its power-law   
spectrum (photon index $\Gamma\sim$1.4, flux corrected for the absorption 
F$_{\rm X}$$\sim$1.2$\times10^{-12}$~erg~cm$^{-2}$~s$^{-1}$) displays a 
prominent iron line (EW=$2.0\pm{0.3}$~keV; Sidoli 2000).

Several bright sources (L$_{\rm X}\sim10^{36}$erg~s$^{-1}$), all previously known,   
have been studied (Sidoli et al. 1999): 
1E~1743.1--2843 (Watson et al. 1981; Cremonesi et al. 1998),  
the   persistent black hole candidates  1E~1740.7--2942 (Hertz \& Grindlay 1984)
and GRS~1758--258 (Sunyaev et al. 1991),
the  X--ray bursters SLX~1744--299, SLX~1744--300, A~1742--294, KS~1741--293 
(Skinner et al. 1990; Pavlinsky, Grebenev \& Sunyaev  1994; Kawai et al. 1988),
and the source  at the GC position
(Watson et al. 1981). 
We also detected two X--ray transients discovered very recently: 
the X--ray burster SAX~J1747.0--2853 (in't Zand et al. 1998; Sidoli et al. 1998) 
and the new superluminal source 
XTE~J1748--288 
(Smith, Levine \& Wood 1998).

The results of the spectroscopy of all these bright sources,  low
mass X--ray binaries containing neutron stars (indicated by type I 
X--ray bursts)
or black holes (with spectra similar to the ``low--hard state" of the 
well know black hole candidate
Cyg~X--1) are shown in Fig.~3, where a single power-law model has been
used to fit the 2--10~keV emission.
The large range in interstellar absorbing column density is also
shown in Fig.~4 in dependence 
of the galactic latitude, that can be translated 
into height from the galactic plane, since all these sources are likely located at the
GC distance.

\begin{figure}[!h]
\vspace{-0.5cm}
 \begin{center}
  \mbox{\epsfig{file=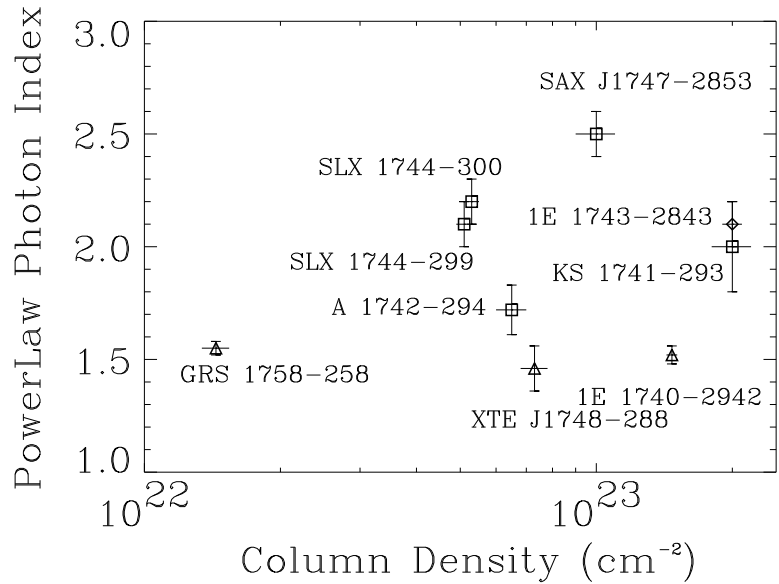,width=12cm}}
  \caption{\em {Absorbing column density and power-law   photon index obtained fitting
  the 2--10~keV spectra of the sources in the GC region with a single power-law. Triangles
  mark the black hole candidates, squares the X--ray bursters}}
 \end{center}
\end{figure}

\begin{figure}[!h]
 \begin{center}
  \mbox{\epsfig{file=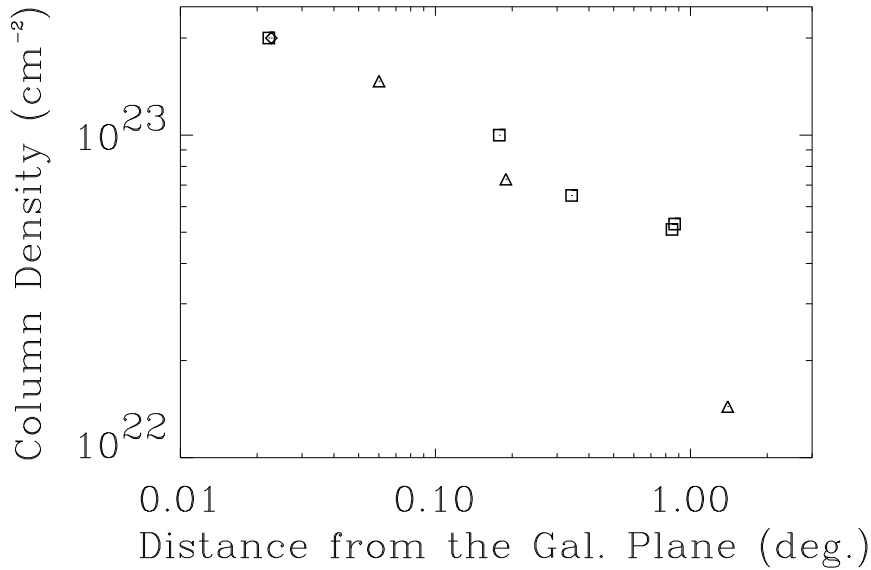,width=12cm}}
  \caption{\em {Distribution  of the absorbing column density (resulting from the spectral
  analysis of  the LMXBs in the GC region) in dependence of   the height from  the 
 galactic plane. The lower column density at $\sim1.3\deg$ from the galactic plane has been
 measured in the direction of GRS~1758--258, located at about $4.5\deg$ from the GC. 
 Symbols have the same meaning as in Fig.~3}}
 \end{center}
\end{figure}

\section {The X--ray Counterpart to Sgr~A*}
 
The large concentration of mass towards the GC direction 
severely complicates the analysis
of the X--ray emission from Sgr~A*, especially with the MECS instrument,
the spatial resolution of which is at a level 
of 1$'$ ($\sim$2.5 pc at the GC distance).
Both the presence of the diffuse emission, peaking at the GC, and 
several  
point--like sources (known from previous missions and displayed in the Fig.~5) 
within few arcminutes from  Sgr~A*, allowed us to place an upper limit to the
X--ray flux contributed by Sgr~A* in the 2--10~keV band.

\vspace{-1cm}
\begin{figure}[!h]
 \begin{center}
  \mbox{\epsfig{file=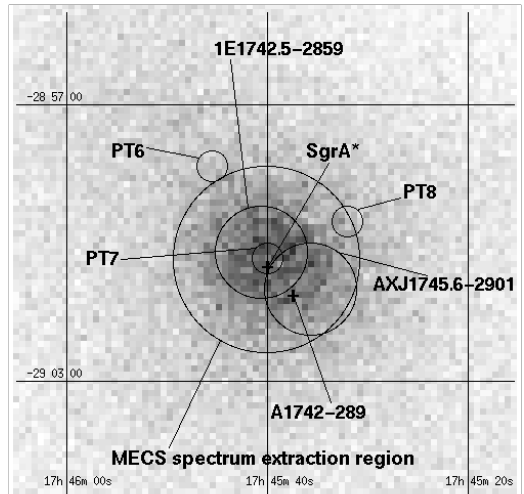,width=7cm}}
  \vspace{-4cm}
\caption{\em {Positions of the X--ray sources in the vicinity 
of Sgr~A* superimposed
on the MECS 2--10~keV image. The error circles have radii of $20''$  
for the ROSAT sources (PT, from Predehl \& Trumper 1994), 
$1'$ for AX~J1745.6--2901 (Maeda et al. 1996)   and $1'$ 
for 1E~1742.5--2859 (Watson et al. 1981). The two crosses mark 
the accurate positions, obtained with   radio observations,
of Sgr~A* (R.A.$=17^{{\rm h}}~45^{{\rm m}}~40.131^{{\rm s}}$, 
Dec.$=-29\deg~00'~27.5''$, Menten et al. 1997) 
and A~1742--289. 
The large circle
($2'$ radius) corresponds to the extraction region 
of the counts used in the spectral analysis. 
All the coordinates are for the J2000 equinox}}
\end{center}
\end{figure}

\vspace{1cm}
With the first imaging of the GC region with the $Einstein$   Observatory 
(Watson et al. 1981, 0.5--4.5 keV) the  point--like source 1E~1742.5--2859,  
positionally coincident with Sgr~A*, was discovered.
A diffuse component around the GC was also detected. 
The  ROSAT satellite observed this same region 
in the 0.1--2.4~keV energy range  (Predehl \& Trumper, 1994)  and,
in addition to the diffuse emission,
showed the presence of three different
sources.
One of these sources   
is  highly absorbed and located within $10''$ from Sgr~A*.
Other X--ray  sources are  located within few arcminutes from Sgr~A*:  
the transient A~1742--289,
discovered in outburst with Ariel~V in 1975 (Branduardi et al. 1976) and the 
LMXB AX~J1745.6--2901 (Maeda et al. 1996).
 
The spectrum and  flux to be ascribed to
the GC point source(s) have been estimated (using the MECS data only) 
extracting counts from a circular region with $2'$ radius 
centered at the position of the X--ray peak flux (see Fig.~5).
A local  background has been taken from an external annular region ($6'-8'$), in order
to subtract the contribution of the  diffuse emission.
Several emission lines (especially the sulfur line at $\sim2.4$~keV and
the iron line at 6.7~keV) are present in this spectrum, indicative  
of a    contamination from the diffuse emission, 
that indeed peaks at the GC position. 
A  good fit has been obtained with a hot plasma model with a temperature 
of $\sim4$~keV, absorbed by a column density, N$_{\rm H}$, of 7$\times10^{22}$~cm$^{-2}$.
No evidence for a 6.4~keV line of
fluorescent origin is present, contrary to what is found from the
surrounding diffuse emission (see next section).
The 2--10~keV flux corrected for the absorption is 
$\sim4\times 10^{-11}$~erg~cm$^{-2}$~s$^{-1}$ and the
luminosity is $\sim3\times 10^{35}$~erg~s$^{-1}$ (a distance of 8.5 kpc is assumed).
Since also the diffuse emission  
contributes to this flux,  only an upper limit to   
the emission from  Sgr~A* can be placed, at a level of  $10^{35}$~erg~s$^{-1}$ (2--10 keV).
This  estimate has been obtained  extrapolating the surface brightness of
the diffuse emission from the surrounding region up to the Sgr~A* position,
and assuming that all the net flux is contributed only by Sgr~A*.
 
Recently, a $Chandra$ observation of the GC detected one weak source, 
probably the  real counterpart to Sgr~A*, with a luminosity, 
L$_{\rm X}$, of $4\times10^{33}$~erg~s$^{-1}$ (0.1--10~keV) (Baganoff et al. 1999).

\section {Diffuse X--ray emission }

The diffuse X--ray emission from the innermost 4 degrees along the 
galactic plane has been mapped with four MECS pointed observations.
The regions considered for the spectral analysis   are indicated 
in   Fig.~6. 
They are marked by circles (radii = 8$'$) representing  
the inner part of the field of view of  four MECS pointings.

\begin{figure}[!h]
 \begin{center}
  \mbox{\epsfig{file=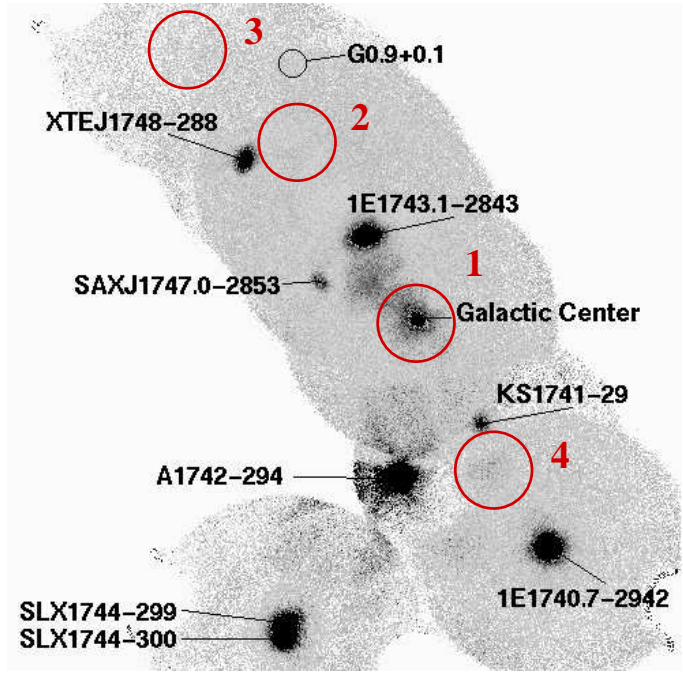,width=9cm}}
\caption{\em {The four circles (radius = 8$'$) in our MECS mosaic 
indicate the  extraction regions for 
the spectral analysis of the diffuse emission.  
They are the inner parts of 4 pointed observations on Sgr~A (1), Sgr~B2 (2),
Sgr~D (3) and Sgr~C (4). Note that fields n.~2 and n.~4 were 
observed twice, but only the observations where  
 the   transient sources XTE~J1748--288 and KS~1741--293 
 were in quiescence have been considered for the spectral analysis
}}
 \end{center}
\label{fig:diff}
\end{figure}

The fields include Sgr~D (l$\sim$1.06, b$\sim$--0.14; field n.~3), 
 Sgr~B2 (l$\sim$0.65, b$\sim$--0.04; field n.~2), 
 Sgr~A (GC, field n.~1) and Sgr~C (l$\sim$359.4, b$\sim$--0.11; field n.~4), 
the main GC molecular clouds complexes as derived with the 
CO surveys (e.g. Oka et al. 1998). 
These molecular clouds are part of a large layer of neutral gas 
distributed along the galactic plane (width$\sim$50~pc, length $\sim$500~pc).
 
The MECS instrument is particularly suited for the study of 
the diffuse sources, due to its simple and unstructured  point spread function. 
All the spectra were extracted from within 8$'$ from 
the pointing direction in order to use the region of the detector 
with the best spatial and spectral properties 
(and also, in some cases, to avoid off--axis point sources). 
Only the spectrum from the Sgr~A pointing was extracted from an 
annular corona with outer and inner radii of 8$'$ and 2$'$ respectively, 
in order to avoid the central point source(s) and the 
probable counterpart to Sgr~A*.

\subsection {Diffuse X--ray emission from the SgrA Complex}

The overall spectrum from the SgrA Complex
displays several emission lines.
A fit with a  thermal bremsstrahlung continuum 
plus three gaussian lines centered at $\sim$1.8, 2.4  and 6.7~keV, 
to account for the   most prominent emission lines in
the spectrum (from Si, S and Fe respectively), resulted in a temperature
of $\sim$13 keV.  
However this temperature 
is too high to be consistent with the presence of the 
low energy emission lines. 
A possible explanation is a multi-temperature plasma. 
Thus, we 
fitted the spectrum with two thermal emission plasma models.  
The best fit parameters are kT$_{\rm 1}$$\sim$0.6~keV
and kT$_{\rm 2}$$\sim$8~keV. 
A gaussian line added at 6.4~keV gives an equivalent width EW$\sim$120~eV. 

To  study the spatial distribution of the diffuse 
emission,  images in two different
energy bands (2--5 keV and 5--10 keV) were extracted.
They  show significantly different morphologies, with the hard emission
with an elliptical shape elongated in the direction of the Galactic plane,
while the soft component shows a triangular shape, spatially correlated with the 
triangular halo of SgrA~East.

The fact that also the spectral data could be well described 
by a two--temperature 
model (with  kT$_{\rm 1}$$\sim$0.6 and kT$_{\rm 2}$$\sim$8~keV),
can suggest that the SgrA~Complex diffuse emission 
may be explained in terms of two plasma components at 
different temperatures and with different spatial distributions. 
In this scenario, it is remarkable that the lower temperature
plasma is well correlated with the   
Sgr~A~East triangular radio halo, a likely evolved SNR (Pedlar et al. 1989).
 
From the spectral parameters of the soft component and the size of the 
SgrA~East halo (radius$\sim$10 pc), 
an electron density n$_{\rm e}$$\sim$3~cm$^{-3}$  
and a pressure P$\sim$3$\times10^{-9}$~erg~cm$^{-3}$ 
can be calculated (Sidoli \& Mereghetti 1999).
This value is consistent with the 
pressure P$_{\rm Sedov}$$\sim$4$\times10^{-9}$~erg~cm$^{-3}$ 
derived  for a SNR  in a Sedov phase. 
In conclusion, if we assume that this
thermal emission is mostly produced by the Sgr~A~East halo, 
its X--ray luminosity, pressure, density, 
temperature of the emitting gas (0.6 keV) 
and size ($\sim$20 pc), match well
with a supernova remnant origin in which thermal line emission
is produced when the expanding supernova
explosion heats the ISM to X--ray temperatures.

\subsection {Diffuse X--ray emission from the Molecular Clouds}

Three fields located on the galactic plane (within about 200 pc from the GC), free
from bright X--ray sources, have been studied (Sidoli 2000). 
The MECS pointings considered here were centered on three molecular clouds
complexes,  Sgr~B2, Sgr~C and Sgr~C (Fig.~6).
The X--ray emission from these fields  is 
harder than that coming from the Sgr~A~Complex, and
displays weaker emission lines at low energy. 
On the other hand the Fe lines around 6--7~keV are 
again particularly bright. An iron line at 6.4~keV is prominent 
in the spectrum from field n.~2, due to fluorescent emission from the
molecular cloud Sgr~B2.

The dependence on the galactic longitude 
of the properties of the iron line diffuse emission are reported in Figs.~7--8.
For comparison, also the parameters for the central 
pointing on Sgr~A (Field n.1) are shown.
The intensity (Fig.~8) of the 6.7~keV iron line 
(in the case of field n.~2 it is a blend of 6.7 and 6.4~keV lines) 
shows a peak at the GC.
Field n.~2 displays an excess with respect to field n.~4, 
located at the same galactocentric
distance, due to the iron emission from Sgr~B2 molecular cloud itself.

\begin{figure}[!h]
\vskip -0.7 truecm
\centerline{\psfig{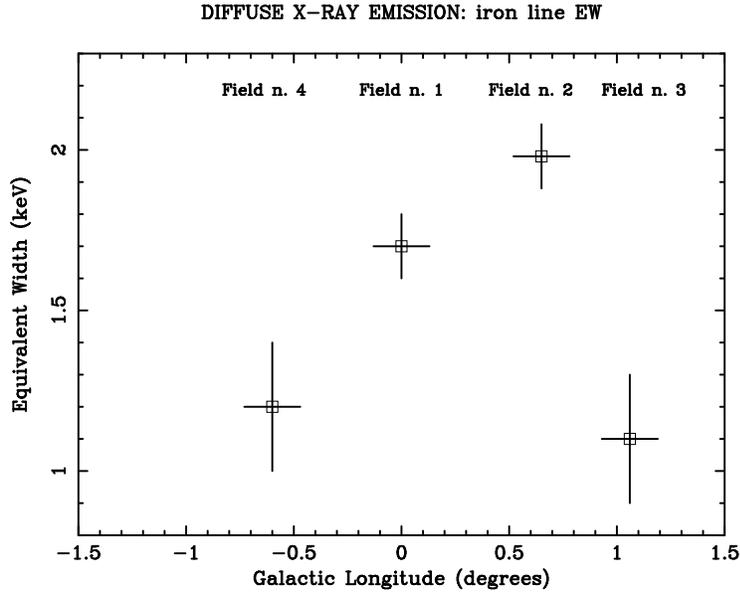}}
\vskip 1.1truecm
\caption{{\small Diffuse X--ray Emission from the GC region: 
equivalent width EW of the iron lines resulting from a fit with a power-law plus a single
gaussian line. The high EW of the emission line contributed by field 2 is due to the 
6.4~keV emission of fluorescent origin contributed by the molecular cloud Sgr~B2  }}
\label{fig:ew}
\end{figure}

\begin{figure}[!h]
\vskip -0.6 truecm
\centerline{\psfig{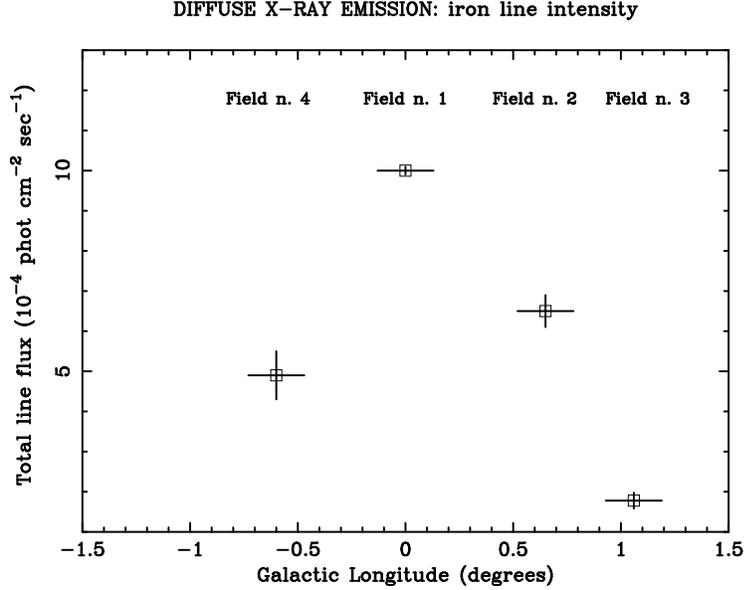}}
\vskip 1.1truecm
\caption{{\small Diffuse X--ray Emission from the GC region:  
intensities of the iron lines resulting from a fit with a power-law plus a single
gaussian line. A clear dependence on the galactic longitude is evident, with the iron line
intensity emitted by the galactic plane peaking at the GC }}
\label{fig:intensity}
\end{figure}

The spectra from the three fields 
have been fit with thermal plasma models, first with single
temperature, then with two--temperature, due to   soft excesses
remaining when fitting with a single component model.
All the results are reported in Table~1. 
The temperatures 
of the soft and hard components from these three fields 
are in the ranges 0.2--0.4~keV  
and 7--9~keV respectively.
 
\vspace{1cm}
\begin{table}[thb]
 \begin{center}
\begin{tabular}[c]{|c|l|l|c|c|} \hline
 Field n.     & N$_{\rm H}$                  & Parameters           &Red.$\chi^2$(dof) &Flux  \\
              &($10^{22}$ cm$^{-2}$)   &                                  &                    &  ($10^{-11}$ erg  cm$^{-2}$ s$^{-1}$)  \\ 
\hline
2             &$11.3^{+0.8}_{-0.9}$     &$T_{\rm 1}=0.23^{+0.02}_{-0.01}$     & 1.29 (284) &     $8.0\pm{0.5}$ \\ 
              &                         &$T_{\rm 2}=7.8^{+1.0}_{-0.5}$        &            &              \\  
              &                         &E$_{\rm line}=6.4$ fixed             &                     &                       \\  
              &                         &$\sigma_{\rm line}=0.15\pm{0.2}$     &                     &                       \\  
              &                         &I$_{\rm line}=2.9\pm{0.4}$           &                     &                       \\  
              &                         &EW$=0.365$                       &                     &                       \\  
\hline
3             &$8\pm{1}$                &$T_{\rm 1}=0.24^{+0.05}_{-0.02}$   & 0.99 (230)          &     $2.8\pm{0.4}$      \\ 
              &                         &$T_{\rm 2}=6.8^{+0.9}_{-0.7}$      &                     &                     \\  
\hline
4             &$10.8\pm{1.3}$           &$T_{\rm 1}=0.35\pm{0.05}$          & 0.96 (152)          &     $9^{+1}_{-2}$      \\ 
              &                         &$T_{\rm 2}=7.5^{+1.3}_{-0.9}$      &                     &                     \\ 
\hline
\end{tabular}
\caption{\em {Results of the spectral analysis of the three fields pointed in the
  direction of the molecular clouds Sgr~B2 (2), Sgr~D(3) and Sgr~C(4). Counts have been
  extracted from within 8 arcmin from the pointing direction;   the 
  cosmic background
  (properly absorbed due to the low galactic latitude)  and the instrumental one 
  have been subtracted.
  The spectra have been fit with a two--temperature  
  plasma model ({\sc mekal}); positive residuals emerging when fitting
  the spectrum from the field~2 require the addition of a gaussian line 
  at 6.4~keV. All energies, EWs  and temperatures are in keV.
  E$_{line}$ is the centroid of the emission line.
The total flux I$_{\rm line}$ from 
the emission line is  in units of 10$^{-4}$~photons~cm$^{-2}$~s$^{-1}$. The hard components
usually contribute about two third of the total luminosity. 
Fluxes (2--10 keV range) have been corrected for the interstellar absorption  
}}
 \end{center} 
\end{table}

The fact that the GC diffuse emission can be well fit 
with a two--temperature plasma model, suggests a comparison with 
the properties of the diffuse emission coming from other regions of the Galactic plane.
In fact, the Galactic plane is permeated by a diffuse X--ray component (the so--called
Galactic Ridge emission) with a two--temperature as well, 
kT$_{\rm 1}$$\sim$0.8~keV and kT$_{\rm 2}$$\sim$8~keV (ASCA observations of the
Scutum Arm region; Kaneda et al. 1997).
 
While the hard components in the Galactic Ridge and in the GC 
display a similar temperature, 
the soft component coming from the GC shows a  significantly
lower temperature (0.2--0.4~keV) with 
respect to the Galactic Ridge emission (0.8~keV). 
 
This discrepancy, other than to different physical 
conditions of the ISM in the GC region,
 could  be due to a  higher interstellar  absorption   that makes 
quite uncertain the estimate of the parameters of 
the soft component in the GC direction. Indeed,
in the Scutum Arm both the less severe interstellar 
absorption and the larger band
of the ASCA instruments (0.5--10 keV) resulted in a 
more precise measurement of the temperature.
Another explanation can be a possible ``contamination" 
of the soft part of the spectra 
by emission physically related with the molecular clouds themselves. 

Molecular clouds can emit X--rays in different ways. 
In several cases the emission is 
produced in the star forming regions naturally located  
inside the clouds (e.g. Koyama et al. 1996b).
Pre--main sequence stars are   strong  X--ray emitters 
(up to 10$^{30}-10^{31}$ erg~s$^{-1}$) with a large variety of behaviours,
both with persistent thermal emission and    with hard flares. 
 
Another possibility can be the emission from 
old isolated neutron stars (ONS) accreting from the dense ISM 
inside the molecular clouds  (Treves \& Colpi 1991; Zane, Turolla \& Treves, 1996).
The spectrum emerging from such emission 
is indeed thermal and rather soft and could explain
the lower soft temperature, with respect to the Galactic Ridge, 
of the fields containing 
molecular clouds. 
In this case, the X--ray luminosity contributed by a single ONS 
depends on the density n$_{\rm cloud}$ of the molecular cloud and on
the relative velocity v of the neutron star with respect to the 
accreting matter: 
L$_{\rm ONS}\sim7\times10^{31}$~n$_{\rm cloud}$~v$_{10}^{-3}$~erg~s$^{-1}$, where v$_{10}$ is the
relative velocity in units of 10~km~s$^{-1}$. 
Assuming v$_{10}\sim$75--100~km~s$^{-1}$ and n$_{\rm cloud}\sim10^{4}$~cm$^{-3}$ for
a typical molecular cloud in the GC region, 
we get L$_{\rm ONS}$$\sim$10$^{32}-10^{33}$~erg~s$^{-1}$. 
The   number of neutron star expected to reside in a single molecular cloud 
can be calculated  from the stellar distribution 
as n$_{\rm ONS}\sim4\times10^{3}~r^{-1.8}$~pc$^{-3}$, where r is the distance from the GC.
Assuming  r$\sim$200~pc and a  cloud volume V$_{\rm cloud}\sim500$~pc$^{3}$,
n$_{\rm ONS}\sim150$~ns/cloud; thus each cloud could contribute at a level of 
about $10^{34}-10^{35}$~erg~s$^{-1}$ to the luminosity of 
the soft component of the diffuse emission
from the GC region.

Molecular clouds can also be the sites of 
reprocessing, scattering and reflection of hard photons  from X--ray sources
located inside or outside the   clouds themselves.  
The strong X--ray emission in the 6.4~keV line from 
Sgr~B2 has be explained by 
Koyama et al. (1996) and by Sunyaev \& Churazov (1996)  with the reflection
of hard X--rays coming from the GC during a past outburst from Sgr~A*.
Also our data require the addition of a 6.4~keV line 
to better fit the spectrum from Sgr~B2,
but   we cannot claim the prevalence of the 6.4~keV fluorescent line 
with respect to the 6.7 iron line as  Murakami et al. (1999).  
 
Observations with a higher spectral and spatial resolution 
are needed in order to precisely define the properties of this 
X--ray emitting cloud and to cast light on the possible link 
with a past high--energy activity of the GC.

\vspace{0.5cm}  
\section{ Acknowledgements}

\begin{small}
The BeppoSAX satellite is a  joint Italian-Dutch programme.
I acknowledge an ESA Fellowship. 
All the results reported here are part of my PhD Thesis,
carried out at the ``G.~Occhialini" Institute of Cosmic Physics (IFC/CNR)
of the C.N.R., Milano (Italy). 

\noindent 
I would like to thank Sandro Mereghetti for his constant support 
and careful supervision. Aldo Treves is  acknowledged for many
suggestions and his guide during the thesis work.
Lucio Chiappetti, Gianluca Israel and Giorgio Matt provided very important  
assistance in the SAX data analysis and it has been a pleasure 
to collaborate with them.
I am very grateful to Silvano Molendi, especially for his help with the analysis 
of the diffuse emission.  
I am grateful to Arvind Parmar for his useful comments on this manuscript.

My special thanks go to Annamaria Borriello, for her important help in numerous occasions and 
for many interesting and constructive discussions.
 
\end{small}

\vspace{2.5cm}
\section { References}

\reff Baganoff, F.,  Angelini, L., Bautz, M., et al.: 1999, AAS {\bf 195}, 6201

\reff Boella, G., Butler, R.C., Perola, G.C., et al.: 1997, A\&AS {\bf 122}, 299

\reff Churazov, E., Gilfanov, M. \& Sunyaev, R.: 1996, ApJ {\bf 464}, L71

\reff Cremonesi, D.I., Mereghetti, S., Sidoli, L. \& Israel, G.L.: 1999, A\&A {\bf 345}, 826

\reff Ghez, A.M., Klein, B.L., Morris, M., Becklin, E.E.: 1998, ApJ {\bf 509}, 678

\reff Hertz, P. \& Grindlay, J.E.: 1984, ApJ, {\bf 278}, 137

\reff in't Zand, J.J. et al.: 1998b, IAU circ. n.6846  

\reff Kaneda, H. et al. 1997, ApJ {\bf 491}, 638

\reff Kawai, N., Fenimore, E.E., Middleditch, J.,et al.: 1988, ApJ {\bf 330}, 130

\reff Koyama, K., et al.: 1989, Nature {\bf 339}, 603

\reff Koyama, K., Maeda, Y., Sonobe, T.,  et al.:  1996, PASJ {\bf 48}, 249

\reff Koyama, K., et al.: 1996b, PASJ 48, L87

\reff Maeda, Y., et al.: 1996, PASJ {\bf 48}, 417 

\reff Menten, K.M., Reid, M.J., Eckart, A. \& Genzel, R.: 1997, ApJ {\bf 475}, L111

\reff Mereghetti, S., Sidoli, L. \& Israel, G.L.: 1998, A\&A {\bf 336}, L81

\reff Murakami, H., Koyama, K., Sakano, M., et al., 2000, ApJ {\bf 534}, 283 

\reff Oka, T., Hasegawa, T., Sato, F., et al., 1998, ApJ Suppl. Ser. {\bf 118}, 455

\reff Pavlinsky, M.N., Grebenev, S.A., \& Sunyaev, R.A.: 1994, ApJ {\bf 425}, 110

\reff Pedlar, A., et al.: 1989, ApJ {\bf 342}, 769

\reff Predehl, P. \& Trumper, J.: 1994, A\&A {\bf 290}, L29

\reff Sidoli, L., Mereghetti, S., Israel, G.L., et al.: 1998, A\&A {\bf 336}, L81

\reff Sidoli, L., Mereghetti, S., Israel, G.L., et al.: 1999, ApJ  {\bf 525}, 215

\reff Sidoli, L. \& Mereghetti, S.: 1999, A\&A {\bf 349}, L49

\reff Sidoli, L.: 2000, PhD Thesis, University of Milan, Italy

\reff Sidoli, L., Mereghetti, S., Israel, G.L., Bocchino, F.: 2000, A\&A submitted

\reff Skinner, G.K., Foster, A.J., Willmore, A.P., Eyles, C.J.: 1990, MNRAS  {\bf 243}, 72

\reff Smith D.A., Levine, A. \& Wood, A.: 1998,  IAU Circ. n.6932

\reff Sunyaev, R.A., Churazov, E., Gilfanov, M., et al.:  1991, A\&A {\bf 247}, L29

\reff Treves, A. \& Colpi, M.: 1991, A\&A {\bf 241}, 107

\reff Valinia, A., et al.: 2000, ApJ in press

\reff Watson, M.G., Willingale, R.,  Grindlay, J.E. \& Hertz, P.: 1981, ApJ {\bf 250}, 142

\reff Zane, S., Turolla, R. \& Treves A.: 1996, ApJ {\bf 471}, 248

\end{document}